\documentclass[12pt]{article}

\usepackage{amsmath,amssymb}
\usepackage{slashed}
\usepackage{xcolor} 
\usepackage{graphicx}
\usepackage{url}
\usepackage{cite}

\def\lsim{\mathrel{\rlap{\lower4pt\hbox{\hskip1pt$\sim$}}
  \raise1pt\hbox{$<$}}}
\def\gsim{\mathrel{\rlap{\lower4pt\hbox{\hskip1pt$\sim$}}
  \raise1pt\hbox{$>$}}}

\newcommand{\beq}{\begin{equation}}
\newcommand{\eeq}{\end{equation}}
\newcommand{\bea}{\begin{eqnarray}}
\newcommand{\eea}{\end{eqnarray}}

\newcommand{\eps}{\epsilon}

\newcommand{\cL}{{\cal L}}
\newcommand{\cM}{{\cal M}}

\graphicspath{{./Figures/}}

\newcommand{\fref}[1]{Fig.~\ref{fig:#1}} 
\newcommand{\eref}[1]{Eq.~\eqref{eq:#1}}

\begin{document}

\vspace*{-2cm}
\begin{flushright}
 LPT Orsay 14-25 \\
\vspace*{2mm}
\today
\end{flushright}

\begin{center}
\vspace*{15mm}

\vspace{1cm}
{\large \bf 
New Observables for CP Violation in Higgs Decays
} \\
\vspace{1.4cm}

{Yi Chen$\,^{a}$,~Adam Falkowski$\,^{b}$,~Ian Low$\,^{c,d}$,\,~Roberto Vega-Morales$\,^{b}$}

 \vspace*{.5cm} 
{\em $^a$Lauritsen Laboratory for High Energy Physics,\\ California Institute of Technology, Pasadena, CA, 92115, USA\\
$^b$Laboratoire de Physique Th\'{e}orique, CNRS - UMR 8627, Universit\'{e} Paris-Sud, Orsay, France\\
$^c$ High Energy Physics Division, Argonne National Laboratory, Argonne, IL 60439, USA\\
$^d$Department of Physics and Astronomy, Northwestern University, Evanston, IL 60208, USA }
\vspace*{.2cm} 

\end{center}

\vspace*{10mm} 
\begin{abstract}\noindent\normalsize
Current experimental data on the 125~GeV Higgs boson still allow room for large CP violation. The observables usually considered in this context are triple product asymmetries, which require an input of four visible particles after imposing momentum conservation. We point out a new class of CP violating observables in Higgs physics  which require only three reconstructed momenta. They may arise if the process involves an interference of  amplitudes with different intermediate particles, which provide distinct ``strong phases'' in the form of the Breit-Wigner widths, in addition to possible ``weak phases'' that arise from CP violating couplings of the Higgs in the Lagrangian. As an example, we propose a forward-backward asymmetry of the charged lepton in the three-body Higgs decay, $h\to \ell^-\ell^+ \gamma$, as a probe for CP-violating Higgs couplings to $Z\gamma$ and $\gamma\gamma$ pairs. Other processes exhibiting this type of CP-violation are also discussed.
\end{abstract}

\vspace*{3mm}

\newpage

\section{Introduction} \label{sec:intro} 
The observation of the 125 GeV Higgs boson at the Large Hadron Collider (LHC)~\cite{:2012gk,:2012gu} marked the beginning of a long-term research program to look for physics beyond the Standard Model (SM) through properties of the Higgs boson.~So far measurements based on the signal strength conform to SM predictions.~However, some properties of the Higgs boson, in particular the tensor structure of its coupling to matter, remain relatively unconstrained by publicly available experimental data.~One particularly interesting possibility is that the Higgs couplings to SM gauge bosons and/or fermions contain new sources of CP-violation. While some of these couplings may be significantly constrained by low-energy precision observables~\cite{McKeen:2012av,Brod:2013cka}, such constraints are not model-independent.~It is therefore important to directly constrain the possibility of CP violating Higgs couplings in high-energy colliders~\cite{Englert:2010ud,Desai:2011yj,Ellis:2012xd,Ellis:2012jv,Djouadi:2013yb,Ellis:2013ywa,Djouadi:2013qya,Englert:2013opa,Godbole:2013saa,Belusca-Maito:2014dpa}. 

There have been many works on direct measurements of CP violation in Higgs physics~\cite{Chang:1993jy,Grzadkowski:1995rx,Gunion:1996vv,Grzadkowski:1999ye,Grzadkowski:2000hm,Plehn:2001nj, Choi:2002jk, Buszello:2002uu,Hankele:2006ma,Godbole:2007cn,Keung:2008ve,Berge:2008dr,Cao:2009ah,DeRujula:2010ys,Gao:2010qx,Berge:2011ij,Bishara:2013vya,Harnik:2013aja,Berge:2013jra,Modak:2013sb,Chen:2014gka}, which all rely on constructing a CP-odd triple product asymmetries.~Such an observable, however, requires presence of three linearly-independent vectors.~Given that the Higgs is a scalar particle and carries no spin, momentum conservation then implies measurements of four visible momenta in order to probe CP violation in the Higgs sector.~One prime example is the azimuthal angle between the two decay planes of a four-body Higgs decay:
\bea
\cos\phi = \frac{(\vec{p}_1\times \vec{p}_2)\cdot (\vec{p}_3\times \vec{p}_4)}{|\vec{p}_1\times \vec{p}_2| \ |\vec{p}_3\times \vec{p}_4|} \ ,
\eea
which appears in channels such as $h\to 4\ell$ and $h \to \tau\tau$.

In general, CP violation occurs through an interference of two amplitudes with different weak phases, that is phases which change sign under a CP transformation.~If, in addition, the amplitudes also contain different strong phases, which do not change sign under CP, then one can construct simpler CP violating observables.~One example is the asymmetry ${\cal A}_{\rm CP}$ of decays into CP conjugate final states $F$ and $\bar F$.~Let us assume that the decay process is described by two interfering amplitudes, ${\cal M}_F ={\cal M}_1 + {\cal M}_2$, which can be written as ${\cal M}_{i} = |c_i| e^{i(\delta_i + \phi_i)}$, where $\delta_i$ and $\phi_i$ are the strong and weak phases, respectively.~This then gives,
\beq
\label{eq:acp}
A_{\rm CP} = {d \Gamma_F - d \Gamma_{\bar F} \over d \Gamma_F + d \Gamma_{\bar F} } \propto |c_1| |c_2| \sin(\delta_1 - \delta_2) \sin(\phi_1 - \phi_2) ,
\eeq 
where we see explicitly that both $\delta_i$ and $\phi_i$ need to be different for the asymmetry to be non-vanishing. 

In flavor physics, where these types of effects have previously been studied, strong phases are often incalculable because they arise from strong interactions.~There are however exceptions when strong phases come from propagation of intermediate state particles.~One well-known example is time evolution of intermediate states that mix with each other, such as the $B^0$--$\overline{B}^0$ system.~Another example that received less attention is strong phases from the propagation of weakly interacting particles with finite widths~\cite{Nowakowski:1991si,Eilam:1991yv,Atwood:1994zm,Bediaga:2009tr, Berger:2011wh}.~In this paper we point out that this latter possibility may arise in the context of decays and associated production of the Higgs boson.~In this case, the weak phases may arise from couplings of the Higgs boson to the SM particles in the Lagrangian, while the strong phases could come from the finite width effects in the Breit-Wigner propagators of intermediate particles. 

There are a number of specific realizations of the above scenario, with applications in both a hadron collider and a lepton collider.~In this paper we focus primarily on the process $h \to \ell^+ \ell^- \gamma$.~In the SM, the $\ell^+\ell^-$ pair could come from an intermediate $Z$ boson or a photon.~We allow the intermediate vector boson to be on or off shell and do not distinguish between them in our notation.~This process can be used to probe the possible CP violating $h\gamma\gamma$ and $hZ\gamma$ couplings.~Similarly one can consider the decay $h \to \ell^+ \ell^- Z$ in which case CP violating $hZ \gamma$, and $hZZ$ couplings are probed.~We will also discuss $f\bar{f} \to Z / \gamma \to h V$, which is related to $h\to 2\ell+V$ by crossing symmetry, and can also be used to probe CP violating $h\gamma\gamma$, $hZ\gamma$ and $hZZ$ couplings.~For all of these cases the strong phase is provided by the width of the $Z$ boson propagating in the intermediate state, while the weak phases may arise from new physics Higgs couplings to matter.

\section {CP Violation in $h\to \ell^-\ell^+ \gamma$ Decays}
\label{sec:hto2lgamma} 

 \begin{figure}[]
 	\begin{center}
	\includegraphics[width=0.3\textwidth]{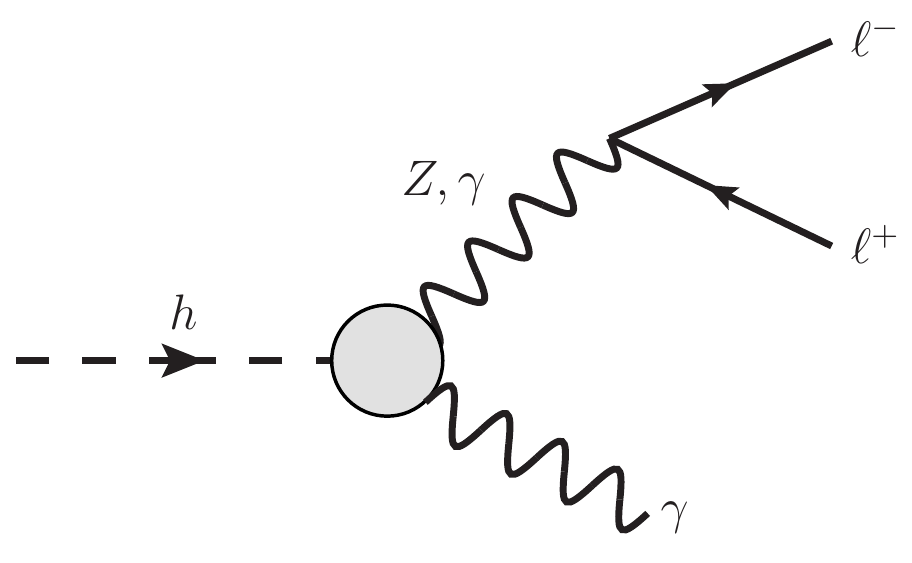}
    \quad
	\vspace*{-2mm}
    \caption{\footnotesize Feynman diagrams for the processes $h \to \ell^-\ell^+ \gamma$ where $\ell = e,\mu$.
    }
    \label{fig:hto2lV}
	\end{center}
\vspace*{-4mm}
\end{figure}

We first focus on the process $h\to \ell^-\ell^+ \gamma$ shown in~\fref{hto2lV}.~The couplings of the Higgs boson to $Z\gamma$ and $\gamma\gamma$ can be parametrized with the following Lagrangian,
\begin{eqnarray}
\label{eq:hcoup}
\mathcal{L} &\supset& \frac{h}{4v} 
\Big( 
2A_{2}^{Z\gamma} F^{\mu\nu}Z_{\mu\nu} + 2A_3^{Z\gamma} F^{\mu\nu} \widetilde{Z}_{\mu\nu} +~A_2^{\gamma\gamma} F^{\mu\nu}F_{\mu\nu} + A_3^{\gamma\gamma} F^{\mu\nu} \widetilde{F}_{\mu\nu} \Big) ,
\end{eqnarray}
where $v=246$ GeV, $V_{\mu\nu} = \partial_\mu V_\nu - \partial_\nu V_\mu$ and $\widetilde{V}_{\mu\nu} = \frac{1}{2} \epsilon_{\mu\nu\rho\sigma} V^{\rho \sigma}$.~We work with effective Higgs couplings for which the SM predicts $A_1^{ZZ} = 2$ at tree level and $A_{2}^{i} \lesssim \mathcal{O}(10^{-2}-10^{-3})$ at 1-loop  ($i = Z\gamma, \gamma\gamma$) .~The $A_{3}^{i}$ are first induced at three loop order~\cite{Korchin:2013ifa} and totally negligible.~We take $A_{2,3}^{i}$ to be momentum independent and real as is done in~\cite{Chen:2012jy,Chen:2013ejz,Chen:2014pia}.~Thus we are neglecting any potential strong phases in the effective couplings, but which in the SM are negligible~\cite{Bishara:2013vya,Czarnecki:1996rx}.~Since the $A_2$ operators are CP-even and $A_3$ are CP-odd,~CP violation must be proportional to products of $A_{2}^i$ and $A_3^j$ in \eref{hcoup}.~In $h\to 4\ell$ we can have CP violation for $i=j$ and $i\neq j$~\cite{Chen:2014gka} because of the ability to form CP-odd triple products from the four visible final state momenta.~As we will see, in the case of the 3-body $h\to \ell^-\ell^+ \gamma$ decay we only obtain CP violation for $i\neq j$ due to the strong phase condition discussed above, i.e.~the Breit-Wigner propagators of the intermediate vector bosons of the interfering amplitudes must be distinct.

To see how CP violation arises in $h\to \ell^-\ell^+ \gamma$ decays it is instructive to analyze the process in terms of helicity amplitudes.~Below we treat the leptons as massless and work in the basis where they have the spin projection $+1/2$ ($R$) or $-1/2$ ($L$) along the direction of motion of $\ell^-$ in the rest frame of the $\ell^-\ell^+$ pair.~We define the $z$-axis by the direction opposite to the motion of photon, which has the polarization tensor $\eps^{\pm 1} = (0,1,\pm i,0)/\sqrt 2$.~The angle $\theta_1$ is then the polar angle of $\ell^-$ in the rest frame of $\ell^+\ell^-$.~Note that for massless leptons, $\ell^+$ and $\ell^-$ must have the same helicity $\lambda_1=\lambda_2\equiv \lambda$, where $\lambda=L,R$.~We denote the helicity amplitudes as $\cM(\lambda,\eps^{\pm 1}) \equiv \lambda_{\pm 1}(\cos \theta_1)$.~In colliders we do not measure helicities, therefore we sum over $\lambda$ and $\epsilon^\pm$ in the amplitude-squared.

Under {\bf P} symmetry all helicities are flipped, while {\bf C} exchanges particles with anti-particle (thus flipping fermion helicities), which corresponds to $\theta_1 \to \pi - \theta_1$.~Thus, the CP transformation relates amplitudes with the same fermion helicity, and opposite photon helicity.~Up to a convention-dependent phase, unbroken CP implies $L_{+1} (\cos \theta_1) = L_{-1}(-\cos \theta_1)$, $R_{+1} (\cos \theta_1) = R_{-1}(-\cos \theta_1)$, in which case,
\beq
\label{eq:helsum}
\sum_{\rm hel.} |\cM|^2 =  |L_{+1}(\cos \theta_1)|^2 + |L_{+1}(-\cos \theta_1)|^2 + |R_{+1}(\cos \theta_1)|^2 + |R_{+1}(-\cos \theta_1)|^2 \ ,
\eeq 
where clearly Eq.~(\ref{eq:helsum}) is symmetric in $\cos \theta_1$.~Therefore a forward-backward asymmetry in the angle $\theta_1$ is a signal of CP violation.~Similarly, unbroken C implies $L_{\pm 1} (\cos \theta_1) = R_{\pm 1}(-\cos \theta_1)$, which implies that the forward-backward asymmetry also requires C violation. 
 
Evaluating the diagram in \fref{hto2lV}, the helicity amplitudes from the intermediate $V = Z, \gamma$ are given by 
\bea
\lambda_{\pm 1} ^V &=& \mp g_{V,\lambda} { (A^{V \gamma}_2 \pm i A_3^{V\gamma}) M_1 (m_h^2 - M_1^2) \over 
 2 \sqrt 2 v (M_1^2 -m_V^2 + i m_V \Gamma_V)} \left (1 \mp \kappa \cos \theta_1 \right ), \quad \lambda=R, L
\eea 
where $\kappa=+1$ for $\lambda=R$ and $-1$ for $\lambda=L$.~We have also defined $M_1$ is the invariant mass of the $\ell^- \ell^+$ pair.~The couplings of the vector boson to left-handed and right-handed leptons are denoted as $g_{V,L}$ and $g_{V,R}$; for the photon we have $g_{V,L} = g_{V,R} = -e$.~In this form we can easily see that the conditions for CP violating asymmetry are satisfied.~More specifically,
\begin{itemize}
\item Two different intermediate particles, $Z$ and $\gamma$, contribute to the same amplitudes.
\item ${\rm Arg} (A_2^{V \gamma} + i A_3^{V\gamma})$,$V=Z,\gamma$, provide different weak phases.
\item ${\rm Arg} (M_1^2 -m_V^2 + i m_V \Gamma_V)$, $V=Z,\gamma$, give distinct strong phases. 
\end{itemize}
It should be clear by now that the forward-backward asymmetry of the $\ell^-$ with respect to the $z$-axis in the $\ell^- \ell^+$ rest frame is a CP-violating observable.
We write the differential decay width as,
\beq
 {d \Gamma \over d M_1^2 d\cos\theta_1} = \left (1 + \cos^2 \theta_1 \right) {d \Gamma_{\rm CPC} \over d M_1^2} + \cos \theta_1 {d \Gamma_{\rm CPV} \over d M_1^2 }. 
\eeq 
The first term is CP conserving and symmetric in $\cos\theta_1$, whereas the second term violates CP and gives rise to the forward-backward asymmetry.~The forward-backward asymmetry can now be computed:
\beq
A_{\rm FB}(M_1) = \frac{\left(\int_0^1 - \int_{-1}^0\right) d\cos\theta_1 {d \Gamma \over d M_1^2 d\cos\theta_1}}{\left(\int_0^1 + \int_{-1}^0\right) d\cos\theta_1d {d \Gamma \over d M_1^2 d\cos\theta_1}} = \frac38\ \frac{d \Gamma_{\rm CPV}/dM_1^2}{ d \Gamma_{\rm CPC}/dM_1^2}\ .
\eeq
Focusing on the CP violating contribution we find,
\beq 
{d \Gamma_{CPV} \over d M_1^2} = 
(A_2^{Z\gamma}A_3^{\gamma\gamma} - A_2^{\gamma\gamma}A_3^{Z\gamma}) \times
{e (g_{Z,R} - g_{Z,L}) m_Z \Gamma_Z (m_h^2 - M_1^2)^3  \over 512 \pi^3 m_h^3 v^2 \left ( (M_1^2 - m_Z^2)^2 + m_Z^2 \Gamma_Z^2 \right )} .
\eeq
The expression is non-zero only in the presence of both CP-even and CP-odd Higgs couplings.~Moreover, we are only sensitive to the products of the Higgs couplings to $Z \gamma$ and $\gamma \gamma$ since this is an interference effect between $Z$ and $\gamma$.~The condition of C violation is provided by the axial coupling of the $Z$ boson to leptons (the Higgs couplings in \eref{hcoup} are C-even), hence the asymmetry is proportional to $(g_{Z,R} - g_{Z,L})$.~The asymmetry vanishes in the limit when $\Gamma_Z$ goes to zero, as then strong phases would be absent.
\begin{figure}[tbh]
	\begin{center}
	\includegraphics[width=0.5\textwidth]{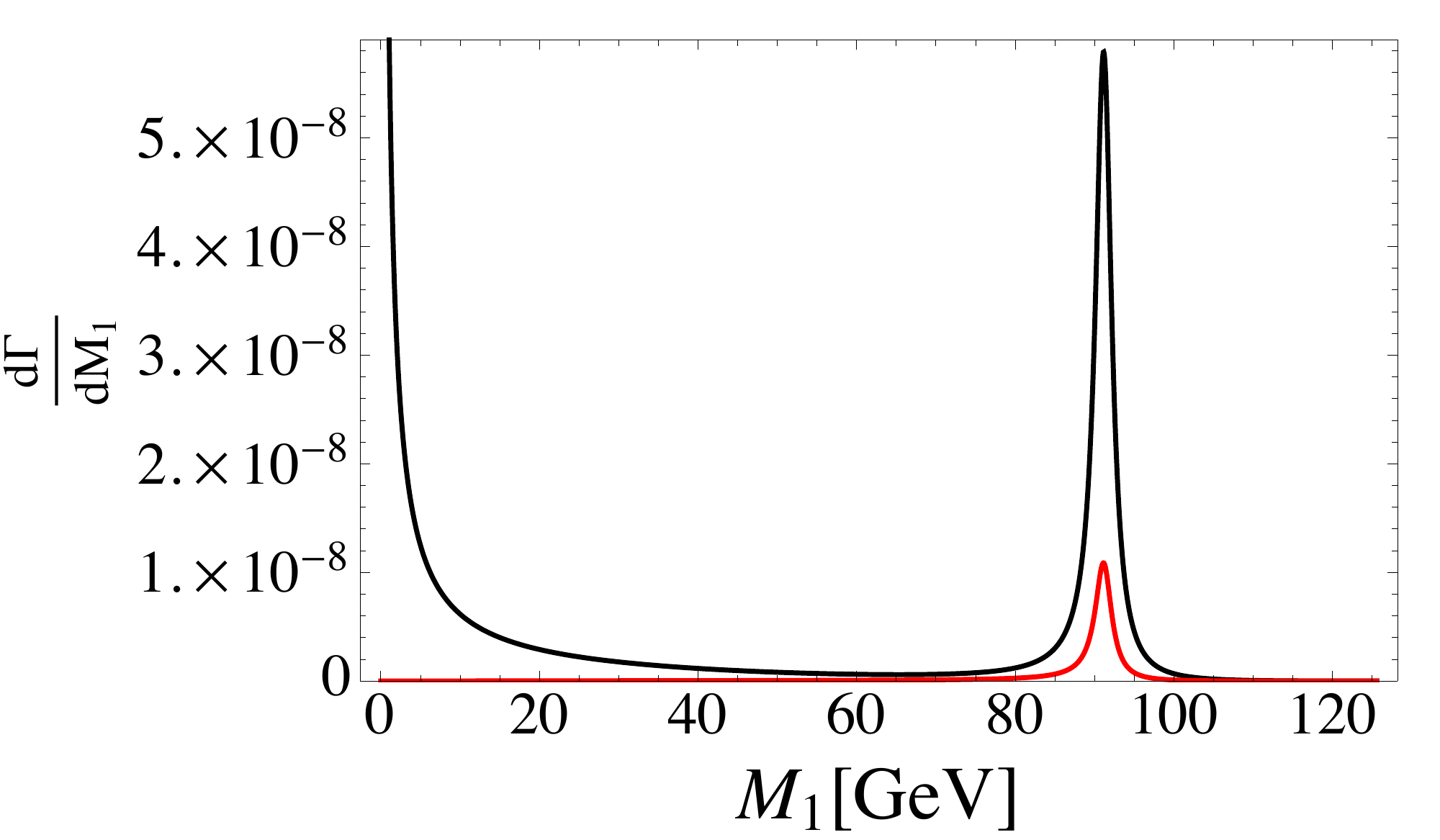}
	\quad
	\includegraphics[width=0.445\textwidth]{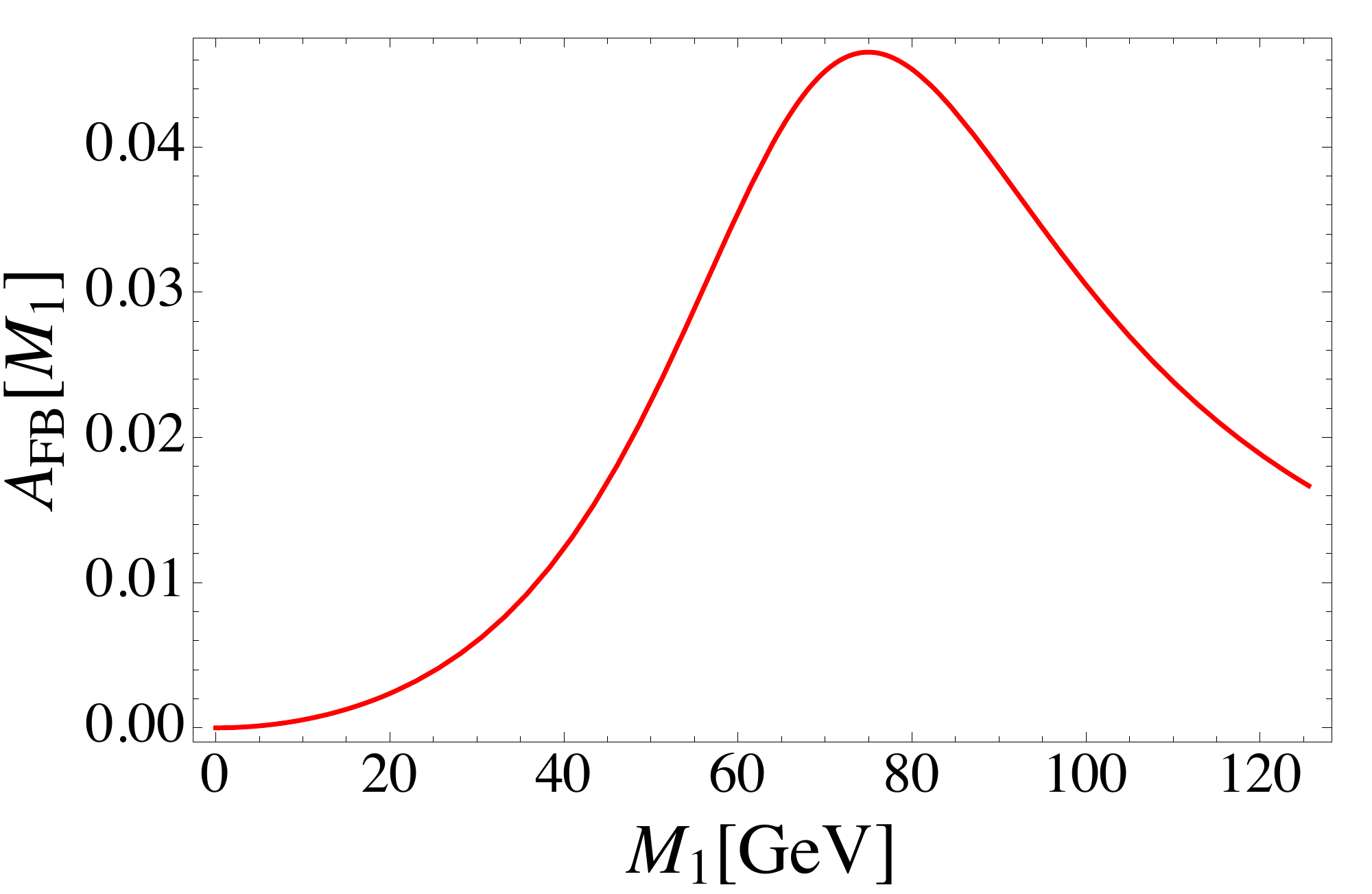}
	\vspace*{-2mm}
     \caption{\footnotesize 
{\bf Left:}~the differential decay rate ${d \Gamma \over d M_1}$ for the symmetric (black) and the asymmetric part~$\times~5$ (red) for $A_3^{Z\gamma} = A_{2\rm SM}^{Z\gamma}$, $A_2^{\gamma\gamma} = A_{2\rm SM}^{\gamma\gamma}$, $A_2^{Z\gamma} = A_3^{\gamma\gamma} =0$.~{\bf Right:}~For the same parameters, the dependence of the signal asymmetry on $M_1$. }
	\label{fig:hzg}
	\end{center}
	\vspace*{-3mm}
\end{figure}
On the left in \fref{hzg} we plot the magnitudes of the symmetric and asymmetric parts of the differential width for a choice of parameters giving rise to SM signal strengths in $\Gamma (h \to Z \gamma )$ and $\Gamma (h \to \gamma \gamma )$.~The shapes of the symmetric and asymmetric parts are very similar on the $Z$ peak.~The rise of the symmetric part for $M_1 \to 0$ is due to the intermediate photon contribution.~On the right in \fref{hzg} we show the differential asymmetry $A_{\rm FB}(M_1)$ for the same choice of parameters.~We can also define the total integrated asymmetry,
\beq
\bar A_{\rm FB} \equiv { 3 \int_{M_0}^{m_h} d M_1\, M_1 {d \Gamma_{\rm CPV} \over dM_1} \over 8 \int_{M_0}^{m_h} d M_1\, M_1 {d \Gamma_{\rm CPC} \over dM_1 } }, 
\eeq 
where the cut $M_1 > M_0$ on the minimum $\ell^- \ell^+$ invariant mass is necessary to cut off the IR divergence due to the intermediate photon.~As long as $M_0$ is not too small, an accurate estimate can be obtained in the narrow width approximation and setting $A_{2,3}^{\gamma \gamma} \to 0$ in the symmetric part.~This way we get,
\beq
\bar A_{\rm FB} \approx 
 { \Gamma_Z \over m_Z}  
 {A_2^{Z\gamma}A_3^{\gamma\gamma} - A_2^{\gamma\gamma}A_3^{Z\gamma} \over (A_2^{Z\gamma})^2+(A_3^{Z\gamma})^2} 
 { 3 e (g_{Z,R} - g_{Z,L}) \over 2 (g_{Z,R}^2 + g_{Z,L}^2)} 
 \approx 0.07 {A_2^{Z\gamma}A_3^{\gamma\gamma} - A_2^{\gamma\gamma}A_3^{Z\gamma} \over (A_2^{Z\gamma})^2+(A_3^{Z\gamma})^2} 
 \eeq 
Clearly, if the CP-odd couplings are of the same order as the CP-even ones, then the only parametric suppression of the asymmetry is by $\Gamma_Z/m_Z \sim 3\%$.~The asymmetry can be larger if $A_2^{Z\gamma}$ is much below the SM value, although that would require a cancellation between the SM $W$ loop and new physics contributions to $h \to Z \gamma$. 

To observe an asymmetry in this channel one must compete not only with the CP conserving part of the $h\to \ell^-\ell^+\gamma$ decay, but also with the much larger irreducible $q\bar{q} \to Z \gamma$ and reducible $Z+X$ (with $X$ faking a photon) backgrounds.~We estimate the expected the significance as follows.~In Ref.~\cite{Gainer:2011aa} it was estimated that after cuts $\sigma_{h} \sim 1.3$~fb for the CP conserving $h\to \ell^-\ell^+\gamma$ decay and $\sigma_{ib} \sim 37$~fb for the irreducible background at $\sqrt{s} =14$~TeV LHC.~We assume here that the reducible background will be of the same order as the irreducible one, thus $\sigma_{b} \sim 2\sigma_{ib}$.~Our signal is $S \sim A_{\rm FB} \sigma_{h} L$, where $L$ is the integrated luminosity, and the background is $B \sim (\sigma_{h} + \sigma_{b})L$.~Then the significance is given by,
\beq
{S \over \sqrt{B}} \sim \left ( A_{\rm FB} \over 0.1 \right ) \sqrt{ L \over 3000~{\rm fb}^{-1}}
\eeq 
This suggests the high-luminosity phase of the LHC would have a chance to observe this asymmetry, especially if a matrix element method analysis similar to what has been done in~\cite{Chen:2012jy,Chen:2013ejz,Chen:2014pia} is used to boost the sensitivity significantly.~This direction is currently under study~\cite{followup2}.

On the other hand, a similar estimate indicates one should be able to probe $A_{\rm FB} \sim 0.05$ in a $100$~TeV $pp$ collider with 3000~fb${}^{-1}$ even using a simpler cut-based approach akin to Ref.~\cite{Gainer:2011aa}. 

\section{CP Violation in Other Processes}
\label{sec:related}

We move to discussing other processes exhibiting this new class of CP violating observables.~In this section we restrict to order of magnitudes estimates of the asymmetry, and briefly comment on the discovery prospects. 

First, we consider the $h\to \ell^-\ell^+ Z$ decay with an on-shell $Z$ boson.~This process is very similar to the $h \to \ell^- \ell^+ \gamma$ decay discussed in the previous section, except that in this case the weak phases may originate from the Higgs couplings to $Z \gamma$ and to $ZZ$.~The former were given in \eref{hcoup} and we parametrize the latter as,
\beq
\cL \supset \frac{h}{4v} 
\Big( A_{1}^{ZZ} Z^{\mu}Z_{\mu} + A_2^{ZZ} Z^{\mu\nu}Z_{\mu\nu} + A_3^{ZZ} Z^{\mu\nu} \widetilde{Z}_{\mu\nu} \Big ). 
\eeq 
The new element here is the tree-level coupling $A_{1}^{ZZ}$ which is expected to be much larger than the loop induced couplings $A_2^i$ and $A_3^i$.~Thus, the $A_{1}^{ZZ}$ squared term will dominate the symmetric CP-conserving part of the differential width, while the interference with $A_3^{Z \gamma}$ will dominate the CP violating part.~Thus, the forward-backward asymmetry parametrically behaves as,
\beq
\label{eq:afb_hllz}
\bar A_{\rm FB}(h \to \ell^- \ell^+ Z) \sim { \Gamma_Z \over m_Z} {A_3^{Z\gamma} \over A_1^{ZZ}} \lesssim 10^{-3} . 
\eeq 
The additional suppression by $A_3^{Z\gamma}/A_1^{ZZ} \sim 10^{-2}$ makes the asymmetry difficult to observe.~Note that the closely related $h\to 4\ell$ process can also probe these tensor structures~\cite{Chen:2014gka}.

The CP violating asymmetry of the kind discussed here may also arise in 2-to-2 scattering of fermions into bosons.~If one can distinguish the incoming and outgoing particle, then one possibility is to define the forward-backward asymmetry with respect to the scattering angle in the center-of-mass frame of the collision.~One example is the process $e^- e^+ \to Z / \gamma \to h Z$ in an electron-positron collider.~At the level of the amplitude, it is related to $h\to \ell^-\ell^+ Z$ by crossing symmetry.~In this case we find,
\beq
\label{eq:afb_eehz}
\bar A_{\rm FB} (e^- e^+ \to h Z) \sim { \Gamma_Z m_Z \over s} {A_3^{Z\gamma} \over A_1^{ZZ}} \lesssim 10^{-4} . 
\eeq 
where $\sqrt{s}$ is the center-of-mass energy of the $e^+ e^-$ collision.~We find the additional suppression factor of $m_Z^2/s$ as compared to \eref{afb_hllz}.~This arises because the amplitude for producing a Higgs boson in association with a transverse $Z$ is parametrically suppressed by $m_Z/\sqrt{s}$ compared to that with a longitudinal $Z$.~Thus, the $h Z$ production cross section is dominated by longitudinal $Z$, which does not give rise to the CP asymmetry.~Due to that suppression, observing the asymmetry requires a large integrated luminosity, well beyond what is expected in the $\sqrt{s}=250$~GeV phase of the ILC.~Furthermore, the asymmetry becomes more difficult to observe as the collision energy is increased.

The same parametric dependence as in \eref{afb_eehz} applies for the process $q \bar q \to Z / \gamma \to h Z$ relevant for hadron colliders.~The additional complication in this case is that the direction of the initial quark vs anti-quark can only be determined statistically, based on the boost of the $h Z$ system in the laboratory frame.~The asymmetry can be larger if the final state $Z$ is replaced with a photon.~For the $f \bar f \to Z / \gamma \to h \gamma$ process, both the symmetric and the asymmetric parts depend only on loop-induced couplings $A_{2,3}^i$.~Moreover, only the transverse polarizations of the final state vector boson are present.~Assuming that the symmetric part is dominated by the intermediate $Z$ exchange, we obtain the same parametric dependence as in the $h \to \ell^- \ell^+ \gamma$ case: 
\beq
\label{eq:afb_eehg}
\bar A_{\rm FB} (f \bar f \to h \gamma) \sim  { \Gamma_Z \over m_Z} 
 {A_2^{Z\gamma}A_3^{\gamma\gamma} - A_2^{\gamma\gamma}A_3^{Z\gamma} \over (A_2^{Z\gamma})^2+(A_3^{Z\gamma})^2} \lesssim 10^{-1} . 
\eeq 
It might be interesting to look into this possibility at a 100 TeV $pp$ collider.

All of the above examples have one common feature:~CP transforms $\cos \theta \to -\cos \theta$, with $\theta$ ($\pi -\theta$) defined by the direction of motion of a fermion $f$ (or anti-fermion $\bar f$) with respect to one of the bosons in the process.~This can be traced to the fact that, while $f$ transforms to $\bar f$, the bosons in these processes are neutral and transform to themselves under CP (up to a helicity flip for vectors).~The consequence is that the forward-backward asymmetry is a CP violating observable.~The situation would be different if {\em both} particle pairs were CP-conjugate.~For example, in the processes $f \bar f \to W^+ W^- $ and $f \bar f \to f' \bar f'$, CP leaves $\theta$ invariant which allows a forward-backward asymmetry to arise without CP violation.

In principle, the asymmetry of the kind discussed here can also be induced by CP violating Higgs couplings to fermions in processes such as $f\bar{f} \to h \to Z \gamma$ (s-channel) interfering with $f\bar{f} \to Z\gamma$ (t-channel).
In practice, however, the asymmetry is suppressed by the fermion mass and by the Higgs width, therefore it is too small to be observable.

\section{Conclusions}
\label{sec:conclution}
In this work we proposed a new class of CP violating observables in Higgs physics without the necessity to construct triple product observables.~These observables can be applied to either three-body decays or 2-to-2 scattering processes involving a Higgs boson at either a hadron or a lepton collider.~They allow measurements of CP violating Higgs couplings to $Z$ and $\gamma$ gauge boson pairs as well as in principle to fermions.~Given that the amount of CP violation in the SM is insufficient to generate the observed baryon asymmetry in the universe and that any observation of CP violation in the Higgs sector would be a sign of physics beyond the SM, searching for these additional sources of CP violation would be of utmost importance in current and future colliders.~We leave a careful study on the sensitivity and reach of this class of observables to future work.

\section*{Acknowledgements}
We thank Roni Harnik, JoAnne Hewett, Joe Lykken, Maria Spiropulu, and Roberto Vega for useful conversations.~We also thank Kunal Kumar for help with validations.~A.F.~and~R.V.M.~are supported by the ERC Advanced Grant Higgs@LHC.~Y.C.~is supported by the Weston Havens Foundation and DOE grant No.~DE-FG02-92-ER-40701.~This work used the Extreme Science and Engineering Discovery Environment (XSEDE), which is supported by National Science Foundation grant number OCI-1053575.~I.L.~is supported in part by the U.S.~Department of Energy under Contracts No.~DE-AC02- 06CH11357 at ANL and No.~DE-SC0010143 at NU.~Three of the authors (A.F., I.L., and R.V.M.) would also like to thank the participants of the workshop ``After the Discovery: Hunting for a Non-Standard Higgs Sector" at Centro de Ciencias de Benasque Pedro Pascual for lively atmosphere and discussions.

\bibliographystyle{JHEP}
\bibliography{CPViolationInHiggsDecays}

\providecommand{\href}[2]{#2}\begingroup\raggedright\begin{thebibliography}{10}

\bibitem{:2012gk}
{\bf ATLAS Collaboration} Collaboration, G.~Aad et~al., {\it {Observation of a
  new particle in the search for the Standard Model Higgs boson with the ATLAS
  detector at the LHC}},  {\em Phys.Lett.} {\bf B716} (2012) 1--29,
  [\href{http://xxx.lanl.gov/abs/1207.7214}{{\tt arXiv:1207.7214}}].

\bibitem{:2012gu}
{\bf CMS Collaboration} Collaboration, S.~Chatrchyan et~al., {\it {Observation
  of a new boson at a mass of 125 GeV with the CMS experiment at the LHC}},
  {\em Phys.Lett.} {\bf B716} (2012) 30--61,
  [\href{http://xxx.lanl.gov/abs/1207.7235}{{\tt arXiv:1207.7235}}].

\bibitem{McKeen:2012av}
D.~McKeen, M.~Pospelov, and A.~Ritz, {\it {Modified Higgs branching ratios
  versus CP and lepton flavor violation}},  {\em Phys.Rev.} {\bf D86} (2012)
  113004, [\href{http://xxx.lanl.gov/abs/1208.4597}{{\tt arXiv:1208.4597}}].

\bibitem{Brod:2013cka}
J.~Brod, U.~Haisch, and J.~Zupan, {\it {Constraints on CP-violating Higgs
  couplings to the third generation}},  {\em JHEP} {\bf 1311} (2013) 180,
  [\href{http://xxx.lanl.gov/abs/1310.1385}{{\tt arXiv:1310.1385}}].

\bibitem{Englert:2010ud}
C.~Englert, C.~Hackstein, and M.~Spannowsky, {\it {Measuring spin and CP from
  semi-hadronic ZZ decays using jet substructure}},  {\em Phys.Rev.} {\bf D82}
  (2010) 114024, [\href{http://xxx.lanl.gov/abs/1010.0676}{{\tt
  arXiv:1010.0676}}].

\bibitem{Desai:2011yj}
N.~Desai, D.~K. Ghosh, and B.~Mukhopadhyaya, {\it {CP-violating HWW couplings
  at the Large Hadron Collider}},  {\em Phys.Rev.} {\bf D83} (2011) 113004,
  [\href{http://xxx.lanl.gov/abs/1104.3327}{{\tt arXiv:1104.3327}}].

\bibitem{Ellis:2012xd}
J.~Ellis, D.~S. Hwang, V.~Sanz, and T.~You, {\it {A Fast Track towards the
  `Higgs' Spin and Parity}},  {\em JHEP} {\bf 1211} (2012) 134,
  [\href{http://xxx.lanl.gov/abs/1208.6002}{{\tt arXiv:1208.6002}}].

\bibitem{Ellis:2012jv}
J.~Ellis, R.~Fok, D.~S. Hwang, V.~Sanz, and T.~You, {\it {Distinguishing
  'Higgs' spin hypotheses using $\gamma \gamma$ and $W W^*$ decays}},  {\em
  Eur.Phys.J.} {\bf C73} (2013) 2488,
  [\href{http://xxx.lanl.gov/abs/1210.5229}{{\tt arXiv:1210.5229}}].

\bibitem{Djouadi:2013yb}
A.~Djouadi, R.~Godbole, B.~Mellado, and K.~Mohan, {\it {Probing the spin-parity
  of the Higgs boson via jet kinematics in vector boson fusion}},  {\em
  Phys.Lett.} {\bf B723} (2013) 307--313,
  [\href{http://xxx.lanl.gov/abs/1301.4965}{{\tt arXiv:1301.4965}}].

\bibitem{Ellis:2013ywa}
J.~Ellis, V.~Sanz, and T.~You, {\it {Associated Production Evidence against
  Higgs Impostors and Anomalous Couplings}},  {\em Eur.Phys.J.} {\bf C73}
  (2013) 2507, [\href{http://xxx.lanl.gov/abs/1303.0208}{{\tt
  arXiv:1303.0208}}].

\bibitem{Djouadi:2013qya}
A.~Djouadi and G.~Moreau, {\it {The couplings of the Higgs boson and its CP
  properties from fits of the signal strengths and their ratios at the 7+8 TeV
  LHC}},  \href{http://xxx.lanl.gov/abs/1303.6591}{{\tt arXiv:1303.6591}}.

\bibitem{Englert:2013opa}
C.~Englert, D.~Goncalves, G.~Nail, and M.~Spannowsky, {\it {The shape of
  spins}},  {\em Phys.Rev.} {\bf D88} (2013) 013016,
  [\href{http://xxx.lanl.gov/abs/1304.0033}{{\tt arXiv:1304.0033}}].

\bibitem{Godbole:2013saa}
R.~Godbole, D.~J. Miller, K.~Mohan, and C.~D. White, {\it {Boosting Higgs CP
  properties via $VH$ Production at the Large Hadron Collider}},  {\em
  Phys.Lett.} {\bf B730} (2014) 275--279,
  [\href{http://xxx.lanl.gov/abs/1306.2573}{{\tt arXiv:1306.2573}}].

\bibitem{Belusca-Maito:2014dpa}
H.~Belusca-Maito, {\it {Effective Higgs Lagrangian and Constraints on Higgs
  Couplings}},  \href{http://xxx.lanl.gov/abs/1404.5343}{{\tt
  arXiv:1404.5343}}.

\bibitem{Chang:1993jy}
D.~Chang, W.-Y. Keung, and I.~Phillips, {\it {CP odd correlation in the decay
  of neutral Higgs boson into Z Z, W+ W-, or t anti-t}},  {\em Phys.Rev.} {\bf
  D48} (1993) 3225--3234, [\href{http://xxx.lanl.gov/abs/hep-ph/9303226}{{\tt
  hep-ph/9303226}}].

\bibitem{Grzadkowski:1995rx}
B.~Grzadkowski and J.~Gunion, {\it {Using decay angle correlations to detect CP
  violation in the neutral Higgs sector}},  {\em Phys.Lett.} {\bf B350} (1995)
  218--224, [\href{http://xxx.lanl.gov/abs/hep-ph/9501339}{{\tt
  hep-ph/9501339}}].

\bibitem{Gunion:1996vv}
J.~F. Gunion, B.~Grzadkowski, and X.-G. He, {\it {Determining the top -
  anti-top and Z Z couplings of a neutral Higgs boson of arbitrary CP nature at
  the NLC}},  {\em Phys.Rev.Lett.} {\bf 77} (1996) 5172--5175,
  [\href{http://xxx.lanl.gov/abs/hep-ph/9605326}{{\tt hep-ph/9605326}}].

\bibitem{Grzadkowski:1999ye}
B.~Grzadkowski, J.~F. Gunion, and J.~Kalinowski, {\it {Finding the CP violating
  Higgs bosons at e+ e- colliders}},  {\em Phys.Rev.} {\bf D60} (1999) 075011,
  [\href{http://xxx.lanl.gov/abs/hep-ph/9902308}{{\tt hep-ph/9902308}}].

\bibitem{Grzadkowski:2000hm}
B.~Grzadkowski, J.~F. Gunion, and J.~Pliszka, {\it {How valuable is
  polarization at a muon collider? A Test case: Determining the CP nature of a
  Higgs boson}},  {\em Nucl.Phys.} {\bf B583} (2000) 49--75,
  [\href{http://xxx.lanl.gov/abs/hep-ph/0003091}{{\tt hep-ph/0003091}}].

\bibitem{Plehn:2001nj}
T.~Plehn, D.~L. Rainwater, and D.~Zeppenfeld, {\it {Determining the structure
  of Higgs couplings at the LHC}},  {\em Phys.Rev.Lett.} {\bf 88} (2002)
  051801, [\href{http://xxx.lanl.gov/abs/hep-ph/0105325}{{\tt
  hep-ph/0105325}}].

\bibitem{Choi:2002jk}
S.~Choi, .~Miller, D.J., M.~Muhlleitner, and P.~Zerwas, {\it {Identifying the
  Higgs spin and parity in decays to Z pairs}},  {\em Phys.Lett.} {\bf B553}
  (2003) 61--71, [\href{http://xxx.lanl.gov/abs/hep-ph/0210077}{{\tt
  hep-ph/0210077}}].

\bibitem{Buszello:2002uu}
C.~Buszello, I.~Fleck, P.~Marquard, and J.~van~der Bij, {\it {Prospective
  analysis of spin- and CP-sensitive variables in H $\to$ Z Z $\to$ l(1)+ l(1)-
  l(2)+ l(2)- at the LHC}},  {\em Eur.Phys.J.} {\bf C32} (2004) 209--219,
  [\href{http://xxx.lanl.gov/abs/hep-ph/0212396}{{\tt hep-ph/0212396}}].

\bibitem{Hankele:2006ma}
V.~Hankele, G.~Klamke, D.~Zeppenfeld, and T.~Figy, {\it {Anomalous Higgs boson
  couplings in vector boson fusion at the CERN LHC}},  {\em Phys.Rev.} {\bf
  D74} (2006) 095001, [\href{http://xxx.lanl.gov/abs/hep-ph/0609075}{{\tt
  hep-ph/0609075}}].

\bibitem{Godbole:2007cn}
R.~M. Godbole, .~Miller, D.J., and M.~M. Muhlleitner, {\it {Aspects of CP
  violation in the H ZZ coupling at the LHC}},  {\em JHEP} {\bf 0712} (2007)
  031, [\href{http://xxx.lanl.gov/abs/0708.0458}{{\tt arXiv:0708.0458}}].

\bibitem{Keung:2008ve}
W.-Y. Keung, I.~Low, and J.~Shu, {\it {Landau-Yang Theorem and Decays of a Z'
  Boson into Two Z Bosons}},  {\em Phys.Rev.Lett.} {\bf 101} (2008) 091802,
  [\href{http://xxx.lanl.gov/abs/0806.2864}{{\tt arXiv:0806.2864}}].

\bibitem{Berge:2008dr}
S.~Berge and W.~Bernreuther, {\it {Determining the CP parity of Higgs bosons at
  the LHC in the tau to 1-prong decay channels}},  {\em Phys.Lett.} {\bf B671}
  (2009) 470--476, [\href{http://xxx.lanl.gov/abs/0812.1910}{{\tt
  arXiv:0812.1910}}].

\bibitem{Cao:2009ah}
Q.-H. Cao, C.~Jackson, W.-Y. Keung, I.~Low, and J.~Shu, {\it {The Higgs
  Mechanism and Loop-induced Decays of a Scalar into Two Z Bosons}},  {\em
  Phys.Rev.} {\bf D81} (2010) 015010,
  [\href{http://xxx.lanl.gov/abs/0911.3398}{{\tt arXiv:0911.3398}}].

\bibitem{DeRujula:2010ys}
A.~De~Rujula, J.~Lykken, M.~Pierini, C.~Rogan, and M.~Spiropulu, {\it {Higgs
  look-alikes at the LHC}},  {\em Phys.Rev.} {\bf D82} (2010) 013003,
  [\href{http://xxx.lanl.gov/abs/1001.5300}{{\tt arXiv:1001.5300}}].

\bibitem{Gao:2010qx}
Y.~Gao, A.~V. Gritsan, Z.~Guo, K.~Melnikov, M.~Schulze, et~al., {\it {Spin
  determination of single-produced resonances at hadron colliders}},  {\em
  Phys.Rev.} {\bf D81} (2010) 075022,
  [\href{http://xxx.lanl.gov/abs/1001.3396}{{\tt arXiv:1001.3396}}].

\bibitem{Berge:2011ij}
S.~Berge, W.~Bernreuther, B.~Niepelt, and H.~Spiesberger, {\it {How to pin down
  the CP quantum numbers of a Higgs boson in its tau decays at the LHC}},  {\em
  Phys.Rev.} {\bf D84} (2011) 116003,
  [\href{http://xxx.lanl.gov/abs/1108.0670}{{\tt arXiv:1108.0670}}].

\bibitem{Bishara:2013vya}
F.~Bishara, Y.~Grossman, R.~Harnik, D.~J. Robinson, J.~Shu, et~al., {\it
  {Probing CP Violation in $h\rightarrow\gamma\gamma$ with Converted Photons}},
   \href{http://xxx.lanl.gov/abs/1312.2955}{{\tt arXiv:1312.2955}}.

\bibitem{Harnik:2013aja}
R.~Harnik, A.~Martin, T.~Okui, R.~Primulando, and F.~Yu, {\it {Measuring CP
  Violation in $h \to \tau^+ \tau^-$ at Colliders}},  {\em Phys.Rev.} {\bf D88}
  (2013) 076009, [\href{http://xxx.lanl.gov/abs/1308.1094}{{\tt
  arXiv:1308.1094}}].

\bibitem{Berge:2013jra}
S.~Berge, W.~Bernreuther, and H.~Spiesberger, {\it {Higgs CP properties using
  the $\tau$ decay modes at the ILC}},  {\em Phys.Lett.} {\bf B727} (2013)
  488--495, [\href{http://xxx.lanl.gov/abs/1308.2674}{{\tt arXiv:1308.2674}}].

\bibitem{Modak:2013sb}
A.~Menon, T.~Modak, D.~Sahoo, R.~Sinha, and H.-Y. Cheng, {\it {Inferring the
  nature of the boson at 125-126 GeV}},  {\em Phys.Rev.} {\bf D89} (2014)
  095021, [\href{http://xxx.lanl.gov/abs/1301.5404}{{\tt arXiv:1301.5404}}].

\bibitem{Chen:2014gka}
Y.~Chen, R.~Harnik, and R.~Vega-Morales, {\it {Probing the Higgs Couplings to
  Photons in $h\rightarrow 4\ell$ at the LHC}},
  \href{http://xxx.lanl.gov/abs/1404.1336}{{\tt arXiv:1404.1336}}.

\bibitem{Nowakowski:1991si}
M.~Nowakowski and A.~Pilaftsis, {\it {CP odd phenomena due to finite width
  effects of the top quark}},  {\em Mod.Phys.Lett.} {\bf A6} (1991) 1933--1942.

\bibitem{Eilam:1991yv}
G.~Eilam, J.~Hewett, and A.~Soni, {\it {CP asymmetries induced by particle
  widths: Application to top quark decays}},  {\em Phys.Rev.Lett.} {\bf 67}
  (1991) 1979--1981.

\bibitem{Atwood:1994zm}
D.~Atwood, G.~Eilam, M.~Gronau, and A.~Soni, {\it {Enhancement of CP violation
  in $B^\pm \to K(i)^\pm$ D0 by resonant effects}},  {\em Phys.Lett.} {\bf
  B341} (1995) 372--378, [\href{http://xxx.lanl.gov/abs/hep-ph/9409229}{{\tt
  hep-ph/9409229}}].

\bibitem{Bediaga:2009tr}
I.~Bediaga, I.~Bigi, A.~Gomes, G.~Guerrer, J.~Miranda, et~al., {\it {On a CP
  anisotropy measurement in the Dalitz plot}},  {\em Phys.Rev.} {\bf D80}
  (2009) 096006, [\href{http://xxx.lanl.gov/abs/0905.4233}{{\tt
  arXiv:0905.4233}}].

\bibitem{Berger:2011wh}
J.~Berger, M.~Blanke, and Y.~Grossman, {\it {A new CP violating observable for
  the LHC}},  {\em JHEP} {\bf 1108} (2011) 033,
  [\href{http://xxx.lanl.gov/abs/1105.0672}{{\tt arXiv:1105.0672}}].

\bibitem{Korchin:2013ifa}
A.~Y. Korchin and V.~A. Kovalchuk, {\it {Polarization effects in the Higgs
  boson decay to $\gamma Z$ and test of $CP$ and $CPT$ symmetries}},  {\em
  Phys.Rev.} {\bf D88} (2013), no.~3 036009,
  [\href{http://xxx.lanl.gov/abs/1303.0365}{{\tt arXiv:1303.0365}}].

\bibitem{Chen:2012jy}
Y.~Chen, N.~Tran, and R.~Vega-Morales, {\it {Scrutinizing the Higgs Signal and
  Background in the $2e2\mu$ Golden Channel}},  {\em JHEP} {\bf 1301} (2013)
  182, [\href{http://xxx.lanl.gov/abs/1211.1959}{{\tt arXiv:1211.1959}}].

\bibitem{Chen:2013ejz}
Y.~Chen and R.~Vega-Morales, {\it {Extracting Effective Higgs Couplings in the
  Golden Channel}},  \href{http://xxx.lanl.gov/abs/1310.2893}{{\tt
  arXiv:1310.2893}}.

\bibitem{Chen:2014pia}
Y.~Chen, E.~Di~Marco, J.~Lykken, M.~Spiropulu, R.~Vega-Morales, et~al., {\it
  {8D Likelihood Effective Higgs Couplings Extraction Framework in the Golden
  Channel}},  \href{http://xxx.lanl.gov/abs/1401.2077}{{\tt arXiv:1401.2077}}.

\bibitem{Czarnecki:1996rx}
A.~Czarnecki and B.~Krause, {\it {On the dipole moments of fermions at two
  loops}},  {\em Acta Phys.Polon.} {\bf B28} (1997) 829--834,
  [\href{http://xxx.lanl.gov/abs/hep-ph/9611299}{{\tt hep-ph/9611299}}].

\bibitem{Gainer:2011aa}
J.~S. Gainer, W.-Y. Keung, I.~Low, and P.~Schwaller, {\it {Looking for a light
  Higgs boson in the overlooked channel}},
  \href{http://xxx.lanl.gov/abs/1112.1405}{{\tt arXiv:1112.1405}}.

\bibitem{followup2}
Y.~Chen, R.~Vega-Morales, et~al. \href{http://xxx.lanl.gov/abs/Work in
  progress}{{\tt Work in progress}}.

\end{thebibliography}\endgroup

\end{document}